\documentclass[prl, twocolumn, 10pt, showpacs, aps, superscriptaddress]{revtex4-1}
\usepackage{amssymb, amsmath, color, graphicx}

\begin{document}

\title{Noise Spectroscopy Using Correlations of Single-Shot Qubit Readout}
\author{T. Fink}
\affiliation{2nd Institute of Physics C, RWTH Aachen University, D-52074 Aachen, Germany}
\affiliation{JARA - Fundamentals of Future Information Technology, D-52425 J\"ulich, Germany}
\author{H. Bluhm}
\affiliation{2nd Institute of Physics C, RWTH Aachen University, D-52074 Aachen, Germany}
\affiliation{JARA - Fundamentals of Future Information Technology, D-52425 J\"ulich, Germany}
\begin{abstract}
A better understanding of the noise causing qubit decoherence is crucial for improving qubit performance. The noise spectrum affecting the qubit may be extracted by measuring dephasing under the application of pulse sequences but requires accurate qubit control and sufficiently long relaxation times, which are not always available. Here, we describe an alternative method to extract the spectrum from correlations of single-shot measurement outcomes of successive free induction decays. This method only requires qubit initialization and readout with a moderate fidelity and also allows independent tuning of both the overall sensitivity and the frequency region over which it is sensitive. Thus, it is possible to maintain a good detection contrast over a very wide frequency range. We discuss using our method for measuring both $1/f$ noise and the fluctuation spectrum of the nuclear bath of GaAs spin qubits.
\end{abstract}


\maketitle

Understanding and reducing decoherence of qubits is of great interest for quantum computing. Pulse sequences like Hahn's spin echo \cite{Hahn1950} and more advanced sequences such as the Carr-Purcell-Meiboom-Gill sequence \cite{Meiboom1958}, concatenated dynamical decoupling \cite{Khodjasteh2007}, or Uhrig dynamical decoupling (UDD) \cite{Uhrig2007} have been demonstrated to decouple a qubit from its noisy environment and therefore reduce decoherence. In addition to enhancing dephasing times, such pulse sequences can also be used for noise spectroscopy by observing the resulting dependence of qubit coherence on the evolution time. Reference \cite{Cywinski2008} shows that, for Gaussian noise, the spectrum's moments can be obtained using UDD. Yuge \textit{et al.} \cite{Yuge2011} proposed a sequence of equidistant $\pi$-pulses to reconstruct the dephasing noise spectrum. A different approach relies on separating the time scales of the noise correlation and extracting long-time correlations via direct measurement while investigating short-time correlations with pulse sequences \cite{Young2012}. However, pulse-sequence-based noise spectroscopy is subject to certain limitations: For fixed pulse sequences, the frequency region over which this technique is sensitive is proportional to the inverse evolution time, but longer durations also increase the total decoherence. This relation makes it hard to probe low-frequency noise: By the time the frequency region of interest is accessible, the qubit is fully dephased, leaving no measurement contrast. Depending on the details of the spectrum, this problem may be to some extent circumvented by adding more pulses and by an appropriate choice of their timing. However, this strategy will eventually be limited by $\pi$-pulse errors and the $T_1$ time of the qubit. Consider, for example, experiments probing the flux noise in superconducting qubits. Direct measurement of the state allows for detection of low-frequency spectral content $\lesssim 1$\,Hz \cite{Bialczak2007,Sank2012} that is limited by the averaging necessary to get a good measurement contrast and may be increased to $100$\,Hz when a single-shot readout is applied \cite{Yan2012}. Carr-Purcell-Meiboom-Gill and UDD sequences have been used to reveal high-frequency content $\gtrsim 0.1$\,MHz \cite{Bylander2011}. Thus, a gap inaccessible with those methods remains. Similar limitations are encountered in GaAs electron spin qubits, where the measured performance of decoupling sequences was likely limited by imperfect control pulses rather than intrinsic noise caused by nuclear spins \cite{Bluhm2011}.

\begin{figure}[t]
\includegraphics{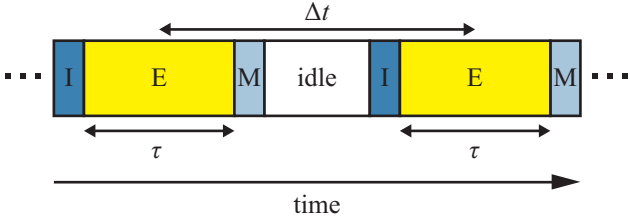}
\caption{Measurement cycle: Qubit initialization (I), evolution (E) for time $\tau$ and measurement (M) of the outcome. The delay time between two evolutions is $\Delta t$. While the initialization and measurement time set a lower limit on $\Delta t$, the idle time can be varied to adjust the delay between successive evolutions.} \label{fig:pulseseq}
\end{figure}

In this Letter, we propose an alternative method for determining the noise spectrum over an extremely wide frequency range without requiring many $\pi$-pulses. It is based on correlating (near) single-shot free induction decay measurements. A typical measurement cycle is depicted in Fig.~\ref{fig:pulseseq} and consists of initializing a $\hat{\sigma}_x$ eigenstate; performing a free evolution under the influence of the noise process for time $\tau$; a projective measurement of the final state; and repeating this process after a delay time $\Delta t$. Hence, depending on the natural preparation and readout axis of a given system, at most two $\pi/2$-pulses per evolution period are required. Averaging over many such measurements allows for computation of the correlation between consecutive measurements as a function of $\Delta t$. A similar method was employed in the experiments of Reilly \textit{et al.} \cite{Reilly2008}, where correlations of temporal averages instead of single-shot readout were used to determine the nuclear spin noise spectrum seen by an electron spin qubit up to $1$\,kHz. A different approach uses dynamical decoupling techniques to introduce an effective delay time and measure correlations of the spin bath of a nitrogen vacancy qubit \cite{DeLange2011}.

After discussing the relationship between noise spectrum and single-shot correlations in general terms, we consider two specific examples from recent experiments to demonstrate the usefulness of our method.

The energy relaxation time of our qubits is assumed to be much longer than the evolution time of interest, and we therefore only consider pure dephasing. This assumption is adequate for many experiments on electron spins \cite{Amasha2008,Petta2005,Reilly2008,Bluhm2010} and for superconducting qubits at operating points where they are linearly sensitive to flux noise \cite{Bylander2011,Nakamura2002,Yoshihara2006,Kakuyanagi2007}. The qubit Hamiltonian therefore is
\begin{equation} \label{eq:hamiltonian}
H=\frac{\hbar}{2} \left[ \Omega + \beta(t) \right] \hat{\sigma}_z,
\end{equation}
where $\Omega$ is the qubit energy splitting and $\beta(t)$ represents a classical Gaussian noise process with $\langle \beta \rangle =0$. We note that Eq.~\ref{eq:hamiltonian} assumes that the noise bath can be described as a classical noise bath. We first set $\Omega=0$ and will discuss the general case later on. The initialized state is a $\hat{\sigma}_x$ eigenstate. A subsequent measurement of $\hat{\sigma}_x$ has an expectation value $P=\cos\left[\Delta\Phi(\tau,t)\right]$, where $\Delta\Phi(\tau,t)=\int_{t-\tau/2}^{t+\tau/2} \beta(t^\prime)\,dt^\prime$ is the phase accumulated during time $\tau$ due to noise. The autocorrelation function of the single-shot outcomes is then given by
\begin{eqnarray} \label{eq:autocorr} \nonumber
\langle P(\tau,t) P(\tau,t+\Delta t) \rangle &=&\frac{1}{2}e^{-\langle \left[ \Delta\Phi(\tau,t)+\Delta\Phi(\tau,t+\Delta t) \right]^2\rangle/2}\\
&+&\frac{1}{2}e^{-\langle \left[ \Delta\Phi(\tau,t)-\Delta\Phi(\tau,t+\Delta t) \right]^2\rangle/2}\\ \nonumber
&=&\frac{1}{2}e^{-\left(\chi_+(\tau,\Delta t)\right) /2}+\frac{1}{2}e^{-\left( \chi_-(\tau,\Delta t) \right) /2},
\end{eqnarray}
and takes values between 1 (perfectly correlated) and 0 (completely independent). We express the exponents $\chi_\pm$ in terms of the noise power spectrum $S_{\beta}(\omega)=\int_{-\infty}^{\infty}e^{i\omega \delta t} \langle\beta(t)\beta(t+\delta t) \rangle d\delta t$ via
\begin{equation} \label{eq:delphi}
\langle \Delta \Phi (\tau,t) \Delta \Phi (\tau,t+\Delta t) \rangle = \int_{0}^{\infty} \frac{d\omega}{\pi} S_\beta (\omega) \frac{F(\omega, \tau, \Delta t)}{\omega^2},
\end{equation}
where $F(\omega, \tau, \Delta t)= 4 \sin^2\left(\frac{\omega \tau}{2}\right) \cos(\omega \Delta t)$ is the filter function analogous to those of decoupling sequences \cite{Cywinski2008}:
\begin{eqnarray} \label{eq:chipm} \nonumber
&\chi_-&(\tau,\Delta t) = 16 \int_0^\infty \frac{d\omega}{\pi}S_\beta(\omega)\frac{\sin^2\left(\frac{\omega\tau}{2}\right) \sin^2\left(\frac{\omega\Delta t}{2}\right)}{\omega^2} \\
&\chi_+&(\tau,\Delta t) = 16 \int_0^\infty \frac{d\omega}{\pi}S_\beta(\omega)\frac{\sin^2\left(\frac{\omega\tau}{2}\right) \cos^2\left(\frac{\omega\Delta t}{2}\right)}{\omega^2}.
\end{eqnarray}
For $\tau \gg T_2^\star$, which is defined by $\langle (\Delta \Phi (T_2^\star,t))^2 \rangle = 1$, the terms $\langle (\Delta \Phi (\tau,t))^2 \rangle$ are much larger than unity but $\chi_-$ can be small because of correlations leading to a partial cancelation of the variances due to the $\langle \Delta \Phi (\tau, t) \Delta \Phi (\tau, t+\Delta t) \rangle$ term. In this case, $\chi_-$ can be measured directly as it is of order unity while $\chi_+$ is negligible because its exponential vanishes. This behavior can also be understood from the filter function: $\chi_+$ has a filter function with larger weight at low frequencies where the spectrum is typically largest (as in the prominent cases of, e.g., $1/f^\alpha$ or Lorentzian-like spectra), whereas the filter function for $\chi_-$ vanishes quadratically for $\omega \to 0$ \footnote{Note that in principle strong anticorrelations can also cause $\chi_+$ to vanish, for example, when the noise is dominated by a narrow range of nonzero frequencies}.

$\chi_-(\tau,\Delta t)$ is the two-parameter generalization of the spin echo decoherence function. This becomes apparent for $\Delta t=\tau$ as the filter function of $\chi_-$ is then identical to that of a spin echo experiment where $F(\omega \tau) = 16\sin^4 \left( 2 \omega \tau / 4 \right)$ and the effect of the reinitialization on $\chi_-$ is thus equivalent to a $\pi$-pulse. Note that initialization and readout need to be negligibly short to reach this limit. While we have assumed $\Omega=0$ so far, a nonzero $\Omega$ adds oscillations $\propto \cos \left(2 \Omega \tau \right)$ to the first term of Eq.~\ref{eq:autocorr}. However, the second term containing $\chi_-$ remains unaffected so that our results remain valid for $\Omega \ne 0$. By varying the delay time $\Delta t$, we can shift the weight of the $\chi_-(\tau,\Delta t)$-filter function in the frequency domain and thus tune the highest sensitivity to the frequency range of interest. Note that, for $\omega \tau \ll 2$, the evolution time $\tau$ controls the gain factor and thus the overall sensitivity. The ability to adjust both the sensitivity's overall magnitude and position mark the strength of our method: Suitable choices of evolution and delay time allow for maintaining a good measurement contrast over a wide frequency range and thus enable the investigation of spectral content. To maximize the range over which our technique may be applied, fast initialization and measurement would be beneficial as they set the lower bound on $\Delta t$. However, low single-shot fidelities can be averaged out and slow initialization can be replaced by projective (nondemolition) measurements.

As a concrete application of our method, we consider electron spin qubits based on gate-defined GaAs quantum dots. For two-electron spin qubits in double quantum dots, the requirements of fast initialization and readout have already been fulfilled \cite{Petta2005,Barthel2009}, and they could likely be achieved with similar techniques for single-spin qubits. For these qubits, the hyperfine interaction couples the electron spins to $\sim 10^6$ nuclei in their respective dots. The resulting Overhauser field $B_{\rm{nuc}}$ contributes to the Zeeman splitting of the electrons, and its temporal fluctuations are the main source of dephasing. For two-electron singlet-triplet qubits, the difference of the Overhauser field in the left and right dot $\Delta B= B_{\rm{nuc}}^{\rm{l}} - B_{\rm{nuc}}^{\rm{r}}$ acts in the same way \cite{Bluhm2011,Taylor2007}.

For this qubit system, $\beta(t) = \frac{g \mu_{\rm{B}}}{\hbar} \Delta B_z(t)$, where $g=-0.44$ is the effective gyromagnetic factor for electrons in GaAs and $\mu_{\rm{B}}$ is Bohr's magneton. The temporal fluctuations in the dots are due to spin diffusion processes that enable distant nuclear spins to exchange polarization. Treating these processes as classical diffusion, one finds that the noise spectrum of the Overhauser fields is proportional to $\omega^{-2}$. It is largely unknown and very interesting to what extent such a description of the spin bath via a classical spectral density, which neglects backaction from the electrons on the nuclei, is valid. Here, we assume that a spectrum is a useful model and focus on how it could be determined. We expect that such a model is most successful at large magnetic fields, where hyperfine-mediated nuclear flip-flops are suppressed. Microscopic theory \cite{Witzel2006} and experiments indicate an $\exp(-t^4)$ decay, which implies that the $\omega^{-2}$ dependence is only valid below $10$\,kHz. However, available data \cite{Bluhm2011,Medford2012} are insufficient to quantitatively characterize the high-frequency behavior. Reference \cite{Biercuk2011} argued that the nuclear spin coupling strength sets an upper cutoff to this behavior and modeled the roll-off as a reduction factor of the form $\exp\left[-\left(\omega/\omega_E\right)^\gamma\right]$. To demonstrate that our method could verify this cutoff behavior, we consider a spectral density of the form 
\begin{equation} \label{eq:spectrum}
S_{\Delta B_z} (\omega)= \frac{S_0}{1+\left(\frac{\omega}{\omega_{L}}\right)^2} \times \exp\left[-\left(\frac{\omega}{\omega_E}\right)^\gamma\right].
\end{equation}
The lower cutoff frequency ($\omega_L$) coarsely mimics the low-frequency behavior of the diffusive dynamics. It ensures a finite rms Overhauser field $\left(\sqrt{\langle \Delta B_z^2 \rangle}\right)$ typically on the order of a few mT \cite{Koppens2005}. The cutoff frequencies, $\omega_L/2\pi$ and $\omega_E/2\pi$, are set to $0.1$\,Hz and $10$\,kHz respectively \cite{Biercuk2011} and $S_0$ determines the overall noise level.

\begin{figure}
\includegraphics{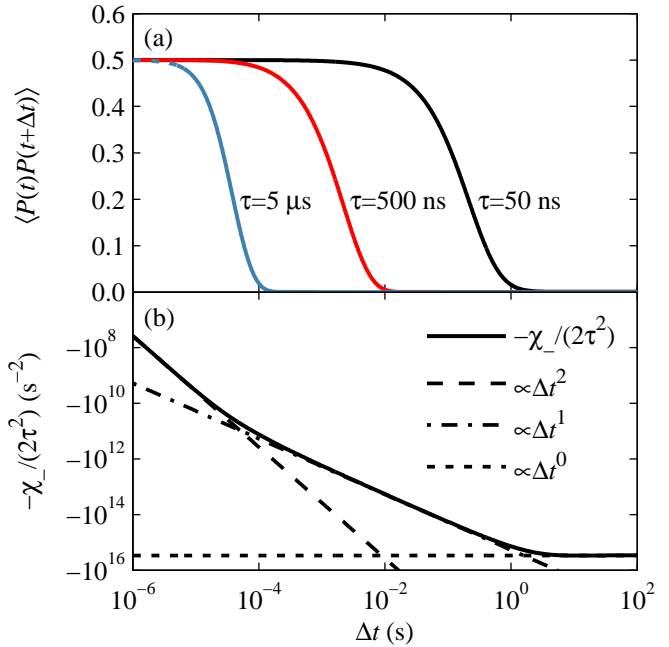}
\caption{(a) Autocorrelation of return probability $P$ for evolution times from $50$\,ns (black) to $5$\,$\mu$s (blue) and exponential cutoff ($\gamma=1$) for the spectrum in Eq.~\ref{eq:spectrum}. Note that, for the dotted part of the blue curve, $\Delta t < \tau$ and thus the correlation function does not exist. (b) Exponent of the dominant term of Eq.~\ref{eq:autocorr}. Note the three different regimes $\Delta t \ll 1/\omega_E$, $1/\omega_E \ll \Delta t \ll 1/\omega_L$, and $1/\omega_L \ll \Delta t$ corresponding to $\Delta t^2$, $\Delta t^1$ and $\Delta t^0$ behavior, respectively, as discussed in the text. Short evolution times $\tau \omega_E \ll 1 $ result in a prefactor $\tau^2$ and show a qualitatively similar behavior.} \label{fig:autocorr}
\end{figure}

The autocorrelation function of the projective measurement is shown in Fig.~\ref{fig:autocorr} (a) and exhibits the expected behavior: For short delay times, the probability that the two measurements return to the initial (singlet) state is perfectly correlated. By increasing $\Delta t$, this correlation decays to zero, where consecutive measurement outcomes are completely independent. For short evolution times, a larger change in $\beta$ is required for decorrelation, so that correlations persist to longer times.

It is useful to consider analytical approximations for the relevant term [$\chi_-$ of Eq.~\ref{eq:autocorr}]. We assume that the evolution time is short compared with the relevant frequencies so that $\tau \omega_E \ll 1$. With this approximation, we obtain
\begin{equation} \label{eq:approximations}
\chi_-(\tau,\Delta t) \approx \begin{cases}
\frac{a}{\pi} c^2 \omega_E \omega_L^2\tau^2S_0\Delta t^2 &\Delta t \ll \omega_E^{-1}\\
c^2 \omega_L^2 \tau^2 S_0 \Delta t & \omega_E^{-1} \ll \Delta t \ll \omega_L^{-1}\\
2c^2 \frac{\tau^2}{\pi} \left(\langle \Delta B_z^2 \rangle \right)^2 & \Delta t  \gg \omega_L^{-1},
\end{cases}
\end{equation}
where $c=\frac{g \mu_{\rm{B}}}{\hbar}$ is a constant and $a=\Gamma \left( \frac{1}{\gamma} + 1 \right)$ is given in terms of the Gamma function $\Gamma(x)$. We find $a=1$ $\left(\sqrt{\pi}/2\right)$ for $\gamma=1$ ($2$). Fig.~\ref{fig:autocorr} (b) compares these approximations (dashed, dash-dotted, and dotted lines) with the exact numerical result (solid line).

We note that, in the regime of short evolution times and small noise variations $\tau^2 \left[ \langle \beta(t)\beta(t+\Delta t) \rangle - \langle \beta^2 \rangle \right] \ll 1$, the single-shot correlation depends linearly on the noise autocorrelation \cite{Reilly2008}, which provides direct access to the noise spectrum. We find that
\begin{equation} \label{eq:linearization}
\langle P(t) P(t+\Delta t) \rangle \approx \frac{1}{2} + \frac{\tau^2}{2} \langle \beta(t) \beta(t+\Delta t) \rangle - \frac{\tau^2}{2} \langle \beta^2 \rangle.
\end{equation}

\begin{figure}
\includegraphics{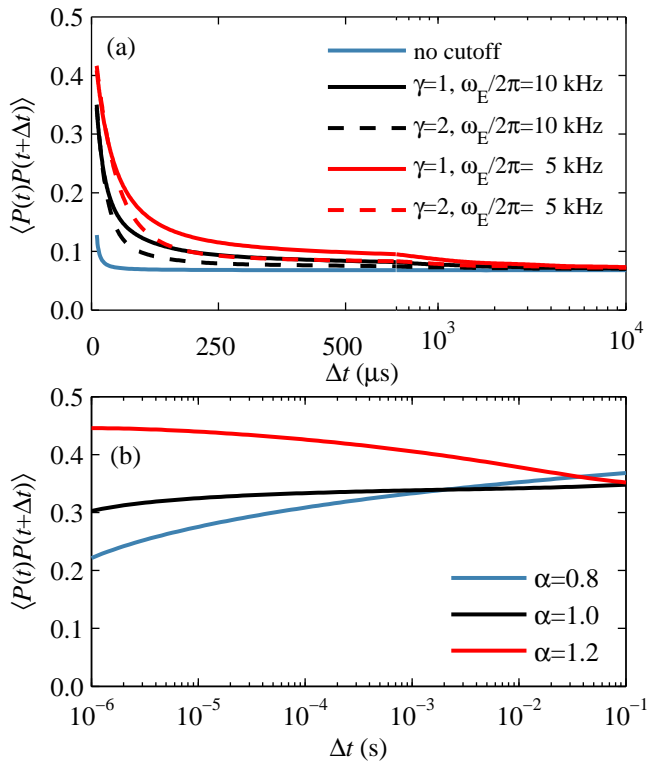}
\caption{(a) Autocorrelation function for the singlet-triplet qubit. Adjusting the evolution time allows for detection of the crossover and slope of curves with different cutoff behavior, thus providing well distinguishable characteristics on an experimentally measurable scale. The lowest (blue) curve shows the autocorrelation in the absence of a cutoff. For $\Delta t>600$\,$\mu$s the scale is semilogarithmic. Note that the plateau at large $\Delta t$ directly reflects the prefactor of the $1/\omega^2$ region of the spectrum. (b) Autocorrelation function for the $1/f^\alpha$ spectrum. For $\tau\ll\Delta t \ll 1/\omega_L$ exact $1/f$ noise leads to a constant result. Deviations from this behavior (curves with negative and positive slope) allow for identification of $\alpha$.} \label{fig:expgauss}
\end{figure}

In order to use our method for noise spectroscopy and, more specifically, to investigate the predicted cutoff behavior, we wish to identify the exponential's power law $\gamma$. For constant $\tau$, a difference between the autocorrelations for exponential and Gaussian cutoff appears for delay times on the order of the inverse cutoff frequency. This frequency also corresponds to the region where the transition between $\Delta t^2$ and $\Delta t$-behavior takes place. For long delay times, the two curves merge as in this region Eq.~\ref{eq:approximations} does not depend on the exact form of the spectrum but only on the rms value of the Overhauser field. The delay time needed to distinguish different cutoff behaviors is thus on the scale of $1 / \omega_E$. Note that in this regime no difference between the logarithm of Eq.~\ref{eq:autocorr} and $-\frac{1}{2} \chi_-$ can be detected, which retrospectively justifies the assumption that the autocorrelation is dominated by $\chi_-$. However, for a constant evolution time, $\chi_-(\tau,\Delta t)$ varies over a large range as a function of $\Delta t$, thus limiting the range of delay times for which the autocorrelation is measurable. We therefore compensate for changes in $\chi_-(\tau, \Delta t)$ due to a varying $\Delta t$ by simultaneously adjusting the evolution time $\tau$. In Fig.~\ref{fig:expgauss} (a) we chose $\tau$ so that the linear term of Eq.~\ref{eq:approximations} always equals 2. The resulting curves allow a clear distinction between different cutoff exponents $\gamma$, while a change of the cutoff frequency $\omega_E$ has a qualitatively different effect. Note that the signal level of $\sim 0.1$ allows a good measurement contrast compared to realistic noise levels over a large range of $\Delta t$.

We now consider using our method for characterization of $1/f^\alpha$ noise which has been found to be an important source of dephasing for several types of superconducting qubits \cite{Nakamura2002,Bialczak2007,Bylander2011,Yoshihara2006}. For a fixed $\tau\ll\Delta t$ and $\alpha>1$, we obtain a decay of the form $\chi_-\propto \Delta t^{(\alpha-1)}$ while for $\alpha=1$, $\chi_-\propto \ln(\Delta t / \tau)\tau^2$. As $\alpha$ is reduced below $1$, the $\Delta t$ dependence of $\chi_-$ becomes increasingly weak, so that a crossover to a constant $\chi_-$ corresponding to white noise is obtained. This behavior reflects the fact that, for noise that is uncorrelated on the scale of $\Delta t$, no correlations and thus no dependence on the delay time will be observed. By adjusting $\tau\propto\Delta t \exp\left[W\left(-\frac{2c}{\Delta t^2}\right)\right]$, exact $1/f$ noise leads to a constant result for $\tau\ll\Delta t$, so that small deviations from $1/f$ in Fig.~\ref{fig:expgauss} can be detected with high sensitivity over a very large frequency range. $W(x)$ is the solution to $x=w\exp(w)$ and $c$ is a constant setting the overall noise level. This allows us in principle to probe arbitrarily low $\omega$ compared to the lower bound of $\sim0.1$\,MHz of Ref. \cite{Bylander2011} which was limited by the $T_2$ time of the qubit. For superconducting qubit systems, initialization is often realized through relaxation, thus making our previous assumption of fast initialization invalid as the initialization time is on the order of $T_1$. To overcome this problem, initialization can be replaced by measurement, which is fast and also provides the information needed to evaluate the autocorrelator.

In conclusion, we have developed a method for noise spectroscopy that fills the frequency gap remaining between direct and pulse-sequence-based noise spectroscopy. Moreover, it enables the identification of different cutoff behaviors and frequency dependencies in spin and superconducting qubits, respectively. As free induction decay pulses and nanosecond control of qubits are well established \cite{Foletti2009,Petta2005,Nowack2007,Koppens2007,Lupascu2006}, experimental implementation should be feasible. Introducing additional operations between the two evolution times might reveal whether the qubit state has non-negligible effects on the noise bath, resulting in different autocorrelation functions.

\acknowledgments{This work was supported by the Alfried Krupp von Bohlen und Halbach Foundation. We would like to thank Lorenza Viola for useful comments.}

\end{document}